\documentclass[aps,pra,showpacs,twoside,twocolumn,10pt]{revtex4-2}
\usepackage[colorlinks=true, citecolor=red, urlcolor=blue ]{hyperref}
\usepackage{epsfig,newlfont,amssymb,amsfonts,amsmath,bm,subfigure,palatino,mathtools,amsthm,braket,times,soul,enumitem,color}
\usepackage[normalem]{ulem}

\usepackage{blindtext}
\usepackage{graphicx}
\usepackage{amsmath}
\usepackage{bm}
\usepackage{hyperref}
\usepackage{geometry}
\usepackage{amsthm}
\usepackage{gensymb}
\usepackage{physics}
\begin{document}
\title{Shared purity and concurrence of a mixture of ground and low-lying excited states as indicators of quantum phase transitions}
\author{George Biswas}
\author{Anindya Biswas}
\affiliation{Department of Physics, National Institute of Technology Sikkim\\ Ravangla, South Sikkim 737 139, India.}
\author{Ujjwal Sen}
\affiliation{Harish-Chandra Research Institute, HBNI, Chhatnag Road, Jhunsi, Prayagraj 211 019, India}

\date{\today}
\begin{abstract}
We investigate the efficacy of shared purity, a measure of quantum correlation that is independent of the separability-entanglement paradigm, as a quantum phase transition indicator in comparison with concurrence, a bipartite entanglement measure. 
The order parameters are investigated for   \textcolor{black}{thermal states, pseudo-thermal states and more,} of the systems considered. In the case of the one-dimensional $J_1-J_2$ Heisenberg quantum spin model and the one-dimensional transverse-field quantum Ising model, shared purity turns out to be as effective as concurrence in indicating quantum phase transitions. In the two-dimensional $J_1-J_2$ Heisenberg quantum spin model, shared purity indicates the two quantum phase transitions present in the model, while concurrence detects only one of them. 
Moreover, we find diverging finite-size scaling exponents for the order parameters near the transitions in odd- and even-sized systems governed by the one-dimensional \(J_1-J_2\) model, as had previously been reported for quantum spins on odd- and even-legged ladders. It is plausible that the divergence is related to a M{\"o}bius strip-like boundary condition required for odd-sized systems, while for even-sized systems, the usual periodic boundary condition is sufficient. 
\end{abstract}
\maketitle
\section{Introduction} \label{I}

Interacting quantum spin Hamiltonians are useful models for studying phenomena in a variety of physical systems. Somewhat recently, it was realized that these Hamiltonians can also be realized in artificial materials, in particular by using ultracold atoms in optical lattices and ion-traps~\cite{doi:10.1080/00018730701223200,Lewenstein2012-jz}
. Realization of these models in artificial materials have, among other things, helped experimenters to reach unprecedented levels of control.

Quantum information and computation is an emerging field at the cross-roads of several fields including physics and information technology. It has recently been realized that quantum information concepts can be utilized to obtain a fresh perspective on many-body phenomena~\cite{RevModPhys.80.517,doi:10.1080/00018730701223200,Lewenstein2012-jz}
. 
The study of quantum phase transitions of spin models using quantum entanglement~~\cite{RevModPhys.81.865,GUHNE20091,doi:https://doi.org/10.1002/9783527805785.ch8,das2017separability} 
 started about two decades ago, when bipartite entanglement was used as an indicator of quantum phase transition in the transverse-field quantum Ising model~\cite{Osterloh2002,PhysRevA.66.032110}. 
Since then, a significant body of work has appeared in the area, connecting quantum critical phenomena of interacting quantum  Hamiltonians with typical concepts that are useful in quantum information, like
bipartite and multipartite entanglement, fidelity, etc.
Along with providing a different perspective and potentially useful understanding of cooperative physical phenomena, such studies are also useful in understanding the potential of a physical substrate for 
realizing information processing tasks.

In this paper, we primarily focus on  frustrated Heisenberg spin models. 
%
%
Quantum phase transitions in the  $J_1-J_2$ Heisenberg quantum spin models in one- and two-dimensional lattices have been investigated using entanglement properties of individual energy eigenstates of the systems~\cite{RECEP,PhysRevA.90.032301,Biswas_2020,PhysRevE.76.061108}. The ground state bipartite as well as genuine multipartite entanglement are unable to conclusively indicate the presence of quantum phase transitions in these spin models~\cite{PhysRevA.90.032301} but  bipartite entanglement of the first excited state indicates their presence~\cite{Biswas_2020}. We approach the same problem from a different perspective. A different observable, namely, shared purity~\cite{PhysRevA.89.032331}, is measured for \textcolor{black}{thermal states, pseudo-thermal states, and for} a\textcolor{black}{n arbitrary} mixture of the ground state and some low-lying excited states of these systems.

The one-dimensional $J_1-J_2$ Heisenberg  quantum spin system lies in the spin fluid phase in the parameter range $\alpha=J_2/J_1 \lesssim 0.24$~\cite{majumdar1969next,PhysRevA.70.052302,PhysRevB.54.9862}. The spin fluid phase is usually observed in frustrated magnets~\cite{doi:10.1126/science.abi8794, Valero2021}. 
The spin model lies in the dimer phase for $\alpha=J_2/J_1 \gtrsim 0.25$~\cite{majumdar1969next,PhysRevA.70.052302,PhysRevB.54.9862}. The increase of the antiferromagnetic next-nearest coupling coefficient $J_2$ enables the formation of dimers~\cite{Xu2021}. We show that shared purity and concurrence (a measure of bipartite entanglement) \textcolor{black}{of the pseudo-thermal state and} a mixture of ground and low-lying excited states detect this spin fluid to dimer quantum phase transition. Moreover, we find that the odd- and even-sized systems have different finite-size scaling exponents.

The two-dimensional  $J_1-J_2$ Heisenberg model lies in the ordinary N\'eel order phase for $\alpha=J_2/J_1 \lesssim 0.40$ and lies in the collinear N\'eel order phase for $\alpha=J_2/J_1 \gtrsim 0.60$. The N\'eel order phase is an antiferromagnetic phase below a sufficiently low temperature (known as the N\'eel temperature)~\cite{PhysRevB.86.144411, refId0}. Above the N\'eel temperature, materials are usually found in the paramagnetic phase. In between $\alpha \approx 0.40$ and  $\alpha \approx 0.60$, there is an intermediate phase, and although there remains some uncertainty, it has been predicted to be a plaquette or columnar dimer phase in the literature~\cite{Schulz_1992,PhysRevB.51.6151,PhysRevB.89.241104,Cysne_2015,PhysRevB.81.144410,Richter2010}.

 Shared purity ($Sp$) is a relatively new quantum correlation measure based on fidelity, which is a measure of the closeness of two quantum states. It is defined as the difference between global and local fidelities~\cite{PhysRevA.89.032331,Liang_2019}, and is given, for a bipartite  quantum state, by
\begin{equation}
Sp=F_{G}-F_{L}.
 \label{Sp_definition}
\end{equation}
The global fidelity ($F_{G}$) calculates the closeness of the quantum state in the argument with all pure states, while the local fidelity ($F_{L}$) calculates the closeness of the same quantum state with pure product states. The computation of the global fidelity is easy, as it is the largest eigenvalue of the density matrix corresponding to the quantum state. However, the optimization over the set of pure product states, required to compute the local fidelity, is often difficult to handle analytically, and may have to be done numerically~\cite{PhysRevA.89.032331}. 
\textcolor{black}{In this paper, to calculate the local fidelity, we maximize the distance of the state under investigation from a set of $10^6$ Haar uniformly generated random pure product states~\cite{doi:10.1063/1.3595693,Biswas_2021}. The convergence of the optimization is checked by using an independent Haar uniform preperation of $10^6$ states. These statements are now added in the revised manuscript.} 
For a pure quantum state, shared purity is the same as the geometric measure of entanglement~\cite{PhysRevA.68.042307}, but it is a different observable for a mixed quantum state~\cite{PhysRevA.89.032331}.

Concurrence is a measure of entanglement, and is usually defined for a  two-qubit quantum state. For a two-qubit density matrix $\rho$, it is defined as 
\begin{equation} \label{e1}
C(\rho)=\max\{0,\lambda_1-\lambda_2-\lambda_3-\lambda_4\},
\end{equation}
where the $\lambda_i\text{'s}$ are  square roots of the eigenvalues of $\rho\tilde{\rho}$ in descending order. Here $\tilde{\rho}$ is the spin-flipped $\rho$ : $\tilde{\rho}=(\sigma_y\otimes\sigma_y)\rho^*(\sigma_y\otimes\sigma_y)$, $\rho^*$ is the complex conjugate of $\rho$ in the computational basis~\cite{PhysRevLett.78.5022,PhysRevLett.80.2245}. 
Concurrence is a monotonically increasing function of the entanglement of formation, which quantifies the necessary amount of singlets to create the bipartite state $\rho$ by local operation and classical   communication~\cite{PhysRevA.53.2046,PhysRevA.54.3824}.

In the upcoming sections~\ref{II},~\ref{III} and~\ref{IV}, we analyze the concurrence and shared purity of \textcolor{black}{thermal states, pseudo thermal states, and} a mixture of ground, and low-lying excited states for the one-dimensional $J_1-J_2$ Heisenberg quantum spin model, the one-dimensional transverse-field quantum Ising model, and the two-dimensional  $J_1-J_2$ Heisenberg quantum spin model respectively. We present a conclusion in Sec.~\ref{V}.

The reason behind the consideration of a mixture of eigenstaes instead of the ground state are that concurrence and shared purity of the ground state can not conclusively detect quantum phase transitions of the one- and two-dimensional $J_1-J_2$ models~\cite{PhysRevA.90.032301}, and it is experimentally difficult to create the ground state of a system. We investigate whether signatures of the quantum phase transitions are still present at low temperatures, wherein the mixing of low-lying eigenstates is inevitable. Therefore, the method employed searches for signatures of quantum phase transitions in the ground state by looking at properties of low-temperature states of the same systems.

\section{One-dimensional antiferromagnetic $J_1-J_2$ Heisenberg model} \label{II}
\par
\begin{figure}[!ht]
\includegraphics[width=\linewidth]{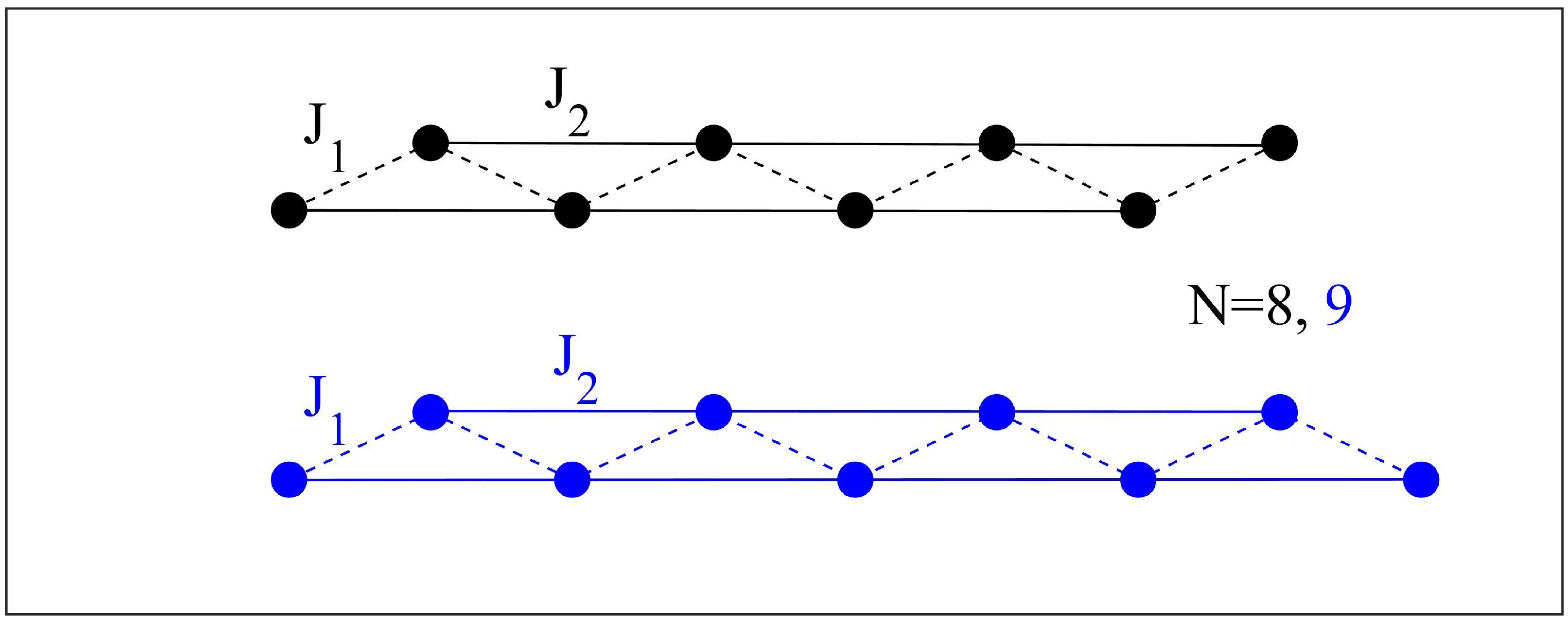}
\caption{The  one-dimensional $J_1-J_2$ Heisenberg quantum spin model imagined as a zig-zag ``chain'' with even and odd number of sites.}
\label{zig_zac}
\end{figure}

 The Hamiltonian for the one-dimensional antiferromagnetic $J_1-J_2$ Heisenberg spin model may be written as
\begin{equation}
H_{1-D}=J_1 \sum_{i=1}^{N}\overrightarrow{\sigma}_i.\overrightarrow{\sigma}_{i+1}+J_2 \sum_{i=1}^{N}\overrightarrow{\sigma}_i.\overrightarrow{\sigma}_{i+2},
 \label{hamiltonian_1d}
\end{equation}
where $\overrightarrow{\sigma}=\sigma^x \hat{x}+\sigma^y \hat{y}+\sigma^z \hat{z}$ with $\sigma^x$, $\sigma^y$ and $\sigma^z$ being the Pauli spin matrices and $N$ is the number of sites in the spin chain. $J_1$ and $J_2$ are both positive and represent the nearest neighbor and the next-nearest neighbor coupling coefficients respectively. The spin system may be imagined as a zig-zag ``chain'' as in Fig.~\ref{zig_zac}, where the dotted lines represent the nearest neighbor interactions and the solid lines represent the next-nearest neighbor interactions. In this sense, the one-dimensional antiferromagnetic $J_1-J_2$ Heisenberg model is 
a quasi- two-dimensional spin model.
In trying to mimic the infinite-size system in a finite-size one, we use periodic boundary conditions, \textit{i.e.}, we assume $\overrightarrow{\sigma}_{N+1}=\overrightarrow{\sigma}_1$. The spin system undergoes a quantum phase transition from the gapless phase to the gapped phase at $\alpha=J_2/J_1\approx0.24.$~\cite{majumdar1969next,PhysRevA.70.052302,PhysRevB.54.9862}. 

In an experiment, it is typically difficult to prepare the ground state of a system of interacting sub-systems \textcolor{black}{and it is already known that the concurrence and shared purity of the ground state alone can not detect quantum phase transitions of the one- and two-dimensional Heisenberg $J_1-J_2$ models~\cite{PhysRevA.90.032301,Biswas_2020}}. 
\textcolor{black}{At some low but non-zero temperature,} it may however be easier to obtain a mixture of the ground state with
low-lying energy levels of the same system.
\textcolor{black}{To investigate along this direction, we analyze four separate cases, as follows. $(i)$ First of all, we consider the thermal state (the canonical equilibrium state) at a particular temperature. $(ii)$ Next, we consider a ``pseudo-thermal" state, where we take the ground state and the first three excited states with Maxwell-Boltzmann probabilities, normalized to unity. $(iii)$ In the third case, we consider a pseudo-thermal state with only the ground and first excited states. $(iv)$ In the final case, we consider a state which is a mixture of the ground state and the first four excited states with exponentially decaying probabilities, normalized to unity. }

\textcolor{black}{We therefore begin with the investigation of detection of quantum phase transition by using concurrence and shared purity of the thermal state for the one-dimensional $J_1-J_2$ Heisenberg quantum spin model. 
A thermal state at a nonzero temperature
may be written as}
\textcolor{black}{\begin{align}\label{thermal_state}
    \rho_{(\beta)}=\frac{1}{Z} \sum_k e^{-\beta E_k}\rho_k,
\end{align}}

\noindent \textcolor{black}{where $E_k$ is the $k^{th}$ eigen energy. And $\rho_k=\frac{1} {d} \sum_{i=1}^d |E_k^i\rangle \langle E_k^i|$ is the equally mixed density matrix of the $k^{th}$ energy level having $d$-fold degeneracy with $|E_k^i\rangle$s being the individual degenerate orthonormal eigenstates spanning that eigenspace. The summation is over all the energy states of the system. The factor in the denominator, $Z=\sum_{k} e^{-\beta E_k}$, is called the partition function and $\beta=\frac{1}{k_B T}$ is proportional to the inverse temperature, where $k_B$ is the Boltzmann constant.}
\begin{figure}[!ht]
\includegraphics[width=\linewidth]{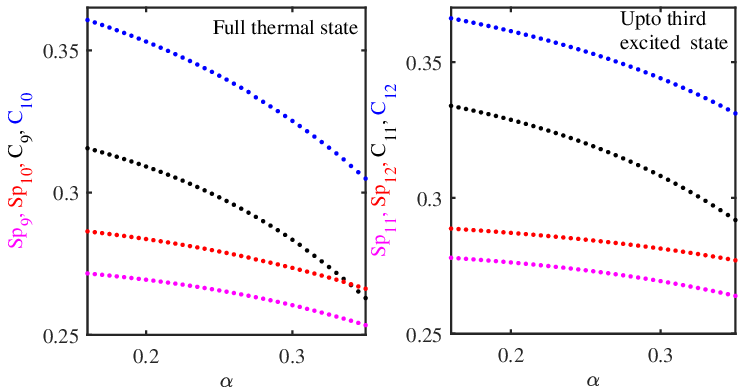}
\caption{\textcolor{black}{Shared purity \(Sp\) and concurrence \(C\) in the one-dimensional \(J_1-J_2\) Heisenberg quantum spin model considering the full thermal state on the left hand side and considering a pseudo-thermal state with contributions considered up to the $3^{rd}$ excited state on the right hand side. The quantities are calculated for nearest neighbor reduced states.
System sizes $N$ are shown in subscripts. When used for concurrence, the vertical axis is in ebits, and it is dimensionless for shared purity. The horizontal axes are dimensionless.}}
\label{thermal}
\end{figure}
\textcolor{black}{We calculate the concurrence and shared purity of the thermal state with $J_1\beta=1$ for the one-dimensional antiferromagnetic $J_1-J_2$ Heisenberg spin model. 
As seen from the left panel of the Fig.~\ref{thermal} the concurrence and shared purity of the thermal state are unable to detect the quantum phase transition.} 

\begin{figure}[!ht]
\includegraphics[width=\linewidth]{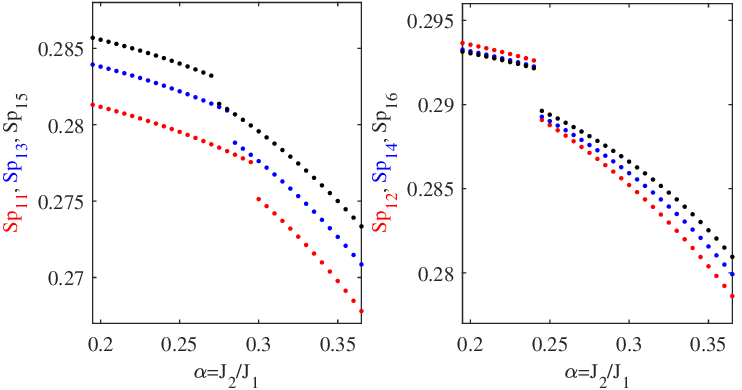}
(a)
\includegraphics[width=\linewidth]{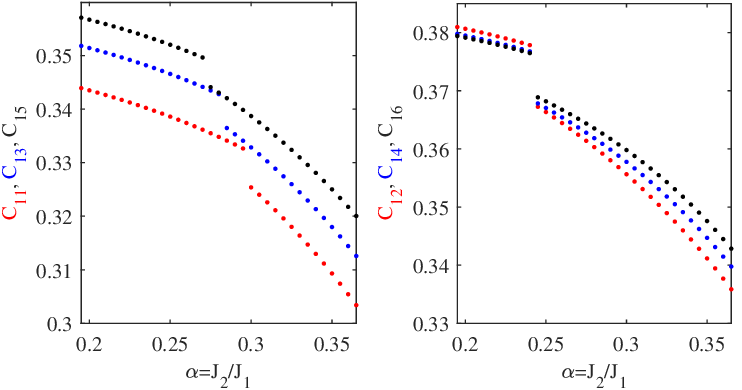}
(b)
\caption{\textcolor{black}{Shared purity and concurrence  in the one-dimensional \(J_1-J_2\) Heisenberg quantum spin model considering pseudo-thermal states with contributions 
up to the $1^{st}$ excited state. The top panels are for shared purity (\(Sp\)), while the bottom ones are for concurrence (\(C\)). The quantities are calculated for nearest neighbor reduced states for different system sizes $N$. The plots on the left are for odd~$N$ (shown in subscript) while the plots on the right are for even~$N$ (shown in subscript). The horizontal axes are dimensionless. The vertical axes on the top panels are also dimensionless, while those for the bottom ones are in ebits.}}
\label{1d_g1}
\end{figure}

\begin{figure}[!ht]
\includegraphics[width=\linewidth]{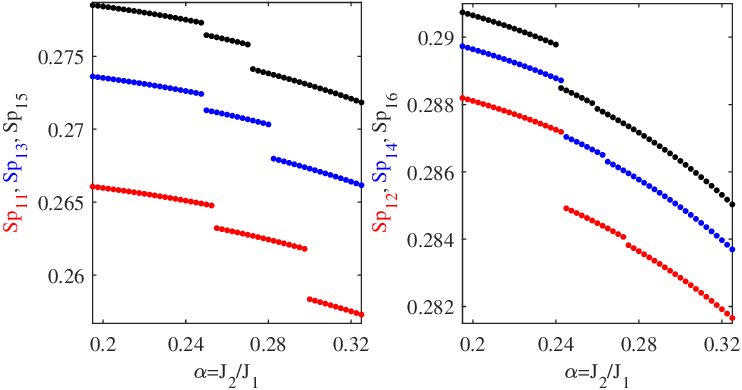}
(a)
\includegraphics[width=\linewidth]{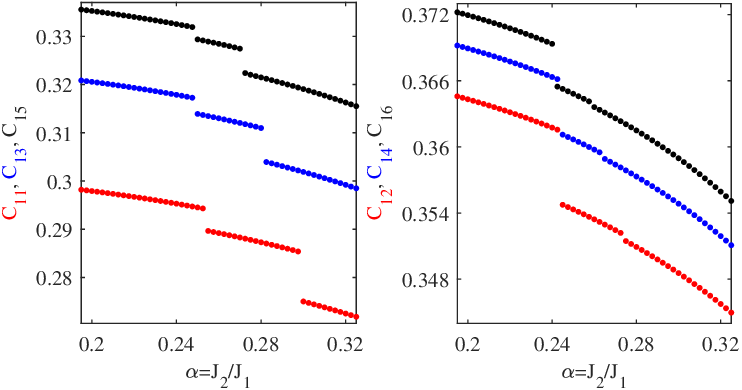}
(b)
\caption{Shared purity and concurrence  in the one-dimensional \(J_1-J_2\) Heisenberg quantum spin model. The top panels  are for shared purity (\(Sp\)), while the bottom ones are for concurrence (\(C\)). The quantities are calculated for nearest neighbor reduced states of the mixture of ground and low-lying excited states of Eq.~(\ref{density_matrix}), 
for different system sizes $N$. 
The plots on the left are for $odd~N$ (shown in subscript) while the plots on the right are for $even~N$ (shown in subscript). The horizontal axes are dimensionless. The vertical axes on the top panels are also dimensionless, while those for the bottom ones are in ebits.}
\label{1d}
\end{figure}

\textcolor{black}{In the right panel of the Fig.~\ref{thermal}, we consider a pseudo-thermal state with contributions up to the third excited state, \textit{i.e.}, the summation over $k$ runs from $0$ to $3$ in the Eq.~(\ref{thermal_state}) and for the expression of the partition function $Z$,
which in this case may be called a pseudo-partition function.  Therefore we observe that if we consider a mixture of a few low-lying states with weights as in the thermal state, we find no indication for the quantum phase transition. The result is shown for both shared purity and concurrence for both odd~$N$ and even~$N$ spin chains. Apparently, the result does not depend on the inverse temperature $\beta$, 
as we do not find any indication of quantum phase transition by varying $J_1\beta$. \textcolor{black}{We denote shared purity and concurrence of system size $N$ by $Sp_N$ and $C_N$ respectively, in the $y$-axis labels of the subfigures of~Fig.~\ref{thermal}, Fig.~\ref{1d_g1} and Fig.~\ref{1d}. }}

\textcolor{black}{ We can, however, detect the quantum phase transition if we consider a mixture of the ground and the first excited states only with weights as in the thermal state, \textit{i.e.}, if summation over $k$ runs from $0$ to $1$ in the Eq.~(\ref{thermal_state}) and in the expression of the pseudo-partition function~$Z$. As seen in the Fig.~\ref{1d_g1}, we observe discontinuities in the value of shared purity and concurrence  indicating the quantum phase transition in the one-dimensional $J_1-J_2$ model.  Therefore, the considered order parameters when applied to the pseudo-thermal state with contributions from the ground and the first excited states only, detect the quantum phase transition.} 

\textcolor{black}{After the consideration of the thermal and pseudo-thermal states,} we have finally considered a mixture of the ground state and low-lying excited states, with exponentially decreasing weights:
\begin{equation}
\rho=\frac{e^{0} \rho_0+e^{-1} \rho_1+e^{-2} \rho_2+e^{-3} \rho_3+e^{-4} \rho_4 + \ldots} {e^{0}+e^{-1}+e^{-2}+e^{-3}+e^{-4} + \ldots}.
 \label{density_matrix}
\end{equation}
The choice of the state is inspired by the canonical equilibrium state of statistical mechanics. 
We restrict our \textcolor{black}{theoretical/mathematical} study up to the $4^{th}$ excited state in Eq.~(\ref{density_matrix}) since higher-order excited states have negligible contributions in the computed values of shared purity and concurrence. \textcolor{black}{that is if we consider higher-order excited states in our calculation, the results will be identical with the one presented.} \textcolor{black}{
Admittedly, it is a very difficult proposition to prepare the mixed state as in Eq.~(\ref{density_matrix}), 
and it is for this reason that we investigated the thermal and pseudo-thermal states beforehand, which may be prepared experimentally with sufficient control over the system's temperature. However, consideration of the mixture in Eq.~\ref{density_matrix} is suitable for quantum phase transition detection in the considered spin systems.
}


We use exact diagonalization techniques and obtain the five lowest eigenvalues and the corresponding eigenstates, and generate the mixed state density matrix as in Eq.~(\ref{density_matrix}). The many-body quantum mixed state is reduced to a nearest-neighbor two-body mixed quantum state, and its concurrence and shared purity are calculated. Fig.~\ref{1d} shows the variation of shared purity and concurrence of the reduced density matrix ($\rho_{nn}$) against the ratio of next-nearest-neighbor and nearest-neighbor coupling coefficients, $\alpha=J_2/J_1$, for different system sizes. 
\par We observe discontinuities in the plotted curves. The positions and  quanta of discontinuities are different for odd~$N$ and even~$N$ spin chains. In the region of interest, there are two discontinuities. The first discontinuities are small for odd~$N$ spin chains and large for even~$N$ spin chains, while the second discontinuities  behave oppositely.
All the discontinuities corresponding to odd~$N$ chains are further away towards the right from the phase transition point, in comparison with those of even~$N$ chains. The positions of these discontinuities are tabulated in Table~\ref{alpha}.

\begin{table}[htbp]
\centering
\begin{tabular}{ |c|c|c|c|c|c|c| } 
 \hline
 $N$ & $\alpha_{c_{near}}^{odd~N}$ & $\alpha_{c_{far}}^{odd~N}$ && $N$ & $\alpha_{c_{near}}^{even~N}$ & $\alpha_{c_{far}}^{even~N}$ \\ 
 \hline
 $9$ & $0.26177$ & $0.33049$ && $8$ & $0.24630$ & $0.31248$ \\ 
 \hline
 $11$ & $0.25401$ & $0.29944$ && $10$ & $0.24449$ & $0.28716$ \\ 
 \hline
 $13$ & $0.25000$ & $0.28243$ && $12$ & $0.24349$ & $0.27285$ \\ 
 \hline
 $15$ & $0.24765$ & $0.27199$ && $14$ & $0.24288$ & $0.26421$ \\ 
 \hline
  & & && $16$ & $0.24248$ & $0.25873$ \\ 
 \hline
\end{tabular}
\caption{Positions on the \( \alpha\)-axis of the near and far discontinuities - near and far from the critical point - for different system sizes, \(N\).
}
\label{alpha}
\end{table}

\textcolor{black}{Note that we observed the discontinuities closer to the phase transition point $(\alpha^{even~N}_{c_{near}})$ for the even-site spin chains and farther from the actual phase transition point  $(\alpha^{odd~N}_{c_{far}})$ for the odd-site spin chains in Fig.~\ref{1d_g1}, where we considered the pseudo-thermal states with a contribution up to the first excited state.}

\par
\begin{figure}[!ht]
\includegraphics[width=\linewidth]{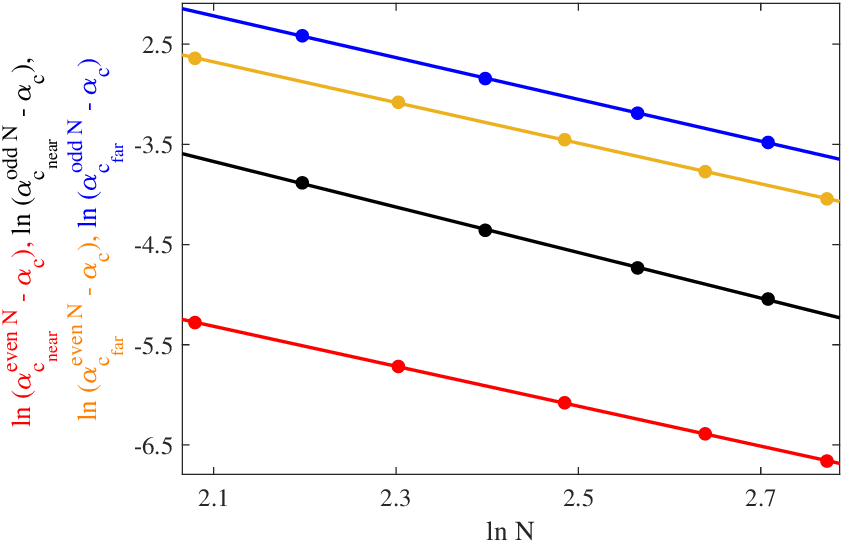}
\caption{Finite-size scaling analysis for the one-dimensional \(J_1-J_2\) Heisenberg quantum spin model. The scaling is separately derived for even- and odd-sized systems, and for the near and the far discontinuities. All quantities are dimensionless. 
}
\label{scalling}
\end{figure}

 It may be noted that the discontinuities in the plots of shared purity versus the driving parameter $\alpha$ coincide with those in the plots of concurrence versus the driving parameter. Asymptotically, the positions of all these discontinuities in the $\alpha$-axis should converge towards the point $\alpha_c\approx0.2412$, where a gapless to gapped quantum phase transition from spin fluid phase to dimer phase exists~\cite{RECEP,PhysRevA.90.032301,Biswas_2020,Tonegawa,OKAMOTO1992433,PhysRevB.54.R9612,PhysRevE.76.061108}. Concurrence of ground state does not indicate this quantum phase transition while that of the $1^{st}$ excited state does indicate it~\cite{Biswas_2020}. It may be noted that the discontinuities in the plot of concurrence versus $\alpha$ for mixed states are nearer to the phase transition point in comparison to the discontinuities in the same plots for the first excited state, for odd-\(N\) spin chains. 
We do the finite-size scaling analysis of all these discontinuities in Fig.~\ref{scalling} and all of them allow straight line fits on a log-log scale, given by the following equation:
\begin{equation}
\ln (\alpha_c^N-\alpha_c)=\beta  \ln N + \mbox{constant}.
 \label{straight_line_fit}
\end{equation}
The slopes of these straight lines are the scaling exponents which tell us how quickly the finite-size critical points converge to the quantum critical point $\alpha_c$ for the corresponding order parameter. The scaling exponents are given by $\beta_{near}^{odd}=-2.271$, $\beta_{far}^{odd}=-2.084$, $\beta_{near}^{even}=-1.993$ and $\beta_{far}^{even}=-2.028$. 

Although all the lines in Fig.~\ref{scalling} reach the same quantum critical point, the scaling exponents are different for odd- and even-site spin chains. Note that the boundary sites of the even spin chains become natural neighbors on  application of the periodic boundary condition, while the boundary sites of the odd spin chains can be properly aligned by forming a M{\"o}bius strip of the quasi two-dimensional spin chain (refer to Fig.~\ref{zig_zac}). One remembers here of different scaling exponents obtained for the generalised geometric measure~\cite{PhysRevA.81.012308}, for odd- and even-legged Heisenberg ladders in~\cite{Roy_2016} (see also~\cite{PhysRevLett.73.886,doi:10.1126/science.271.5249.618,PhysRevB.45.5744,PhysRevB.49.8901,Rice_1993}
). The spin chains investigated here are much simpler systems manifesting a similar phenomenon.

\begin{figure}[!ht]
\includegraphics[width=\linewidth]{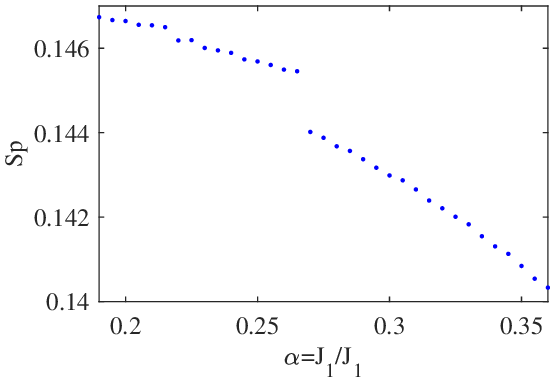}
\caption{\textcolor{black}{Shared purity of three contiguous spins in the one-dimensional $J_1-J_2$ Heisenberg quantum spin model considering the pseudo-thermal state with the ground and first excited states. The vertical axis represents the shared purity, while the horizontal axis represents $\alpha=J_2/J_1$. Both axes are dimensionless.}}
\label{x}
\end{figure}

\textcolor{black}{Until now, we have countered shared purity for two-site reduced states only. Shared purity is however well-defined for density operators of an arbitrary number of sites also. We now analyze the status of shared purity as an order parameter for detecting the quantum phase transition in the one-dimensional $J_1-J_2$ model for reduced density operator of three contiguous spins for the state formed by mixing the ground state with the first excited state. The results are depicted in Fig.~\ref{x}, and we see that just like the two-site shared purity, the three-site quantity can also successfully signal the existence of the critical point of the model.}

\section{One-dimensional transverse-field quantum Ising model} \label{III}
The Hamiltonian of the transverse-field quantum Ising model can be written as 
\begin{equation}
H_{IS}=\lambda \sum_{i=1}^N {\sigma}_i^x{\sigma}_{i+1}^x+ \sum_{i=1}^N {\sigma}_i^z,
 \label{hamiltonian_iz}
\end{equation}
where $\sigma^x$ and $\sigma^z$ are Pauli spin matrices and $\lambda$ is the ratio of the $xx$ coupling constant to  the transverse magnetic field strength.
We now discuss the concurrence of a mixture of the ground and low-lying excited states of the transverse-field quantum Ising model, whose ground state concurrence successfully detects the quantum phase transition present in the system~\cite{Osterloh2002}. For $\lambda<1$ and $\lambda>1$ the spin system lies in gapped phases.
But at $\lambda=1$ the spin model has gapless excitations, and the system  undergoes a quantum phase transition~\cite{Osterloh2002,Maharaj13430}.

\begin{figure}[!ht]
\includegraphics[width=\linewidth]{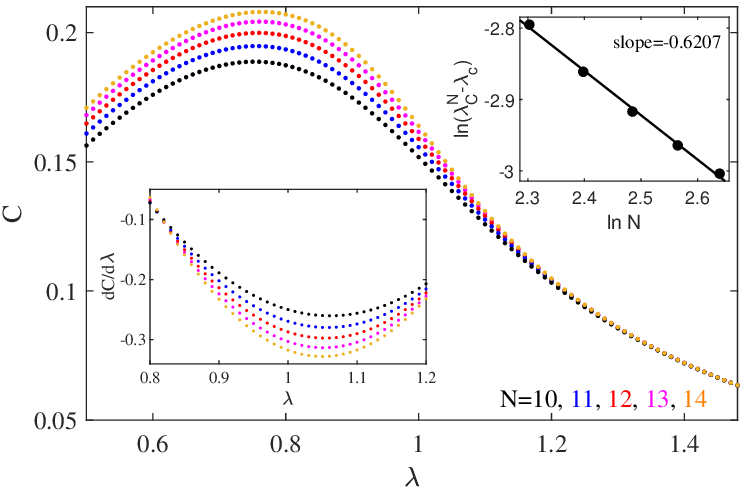}
\caption{\textcolor{black}{We analyze the concurrence of a mixture of the ground and the first excited states with weights as in the thermal state of the one-dimensional transverse-field quantum Ising model for different system sizes and their derivatives (in lower inset), and the finite-size scaling of the minimum points of the derivatives (in upper inset).. The colors for the different system sizes are the same as in the legend.  All horizontal axes are dimensionless. The vertical axis of the upper inset is also dimensionless, while those of the main plot and the lower inset are in ebits.}}
\label{iz_thermal_shortest_con}
\end{figure}

\begin{figure}[!ht]
\includegraphics[width=\linewidth]{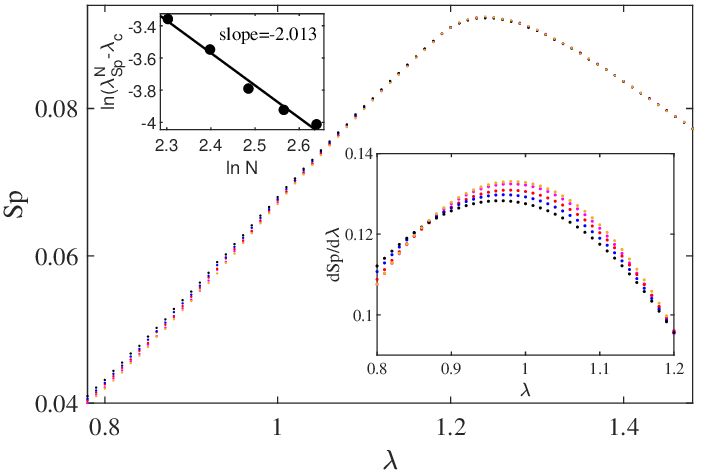}
\caption{\textcolor{black}{We analyze here the shared purity of a mixture of the ground and the first excited states with weights as in the thermal state of the one-dimensional transverse-field quantum Ising model for different system sizes. The derivatives are plotted in the lower inset, while the scaling analysis of the maximum points of the derivatives is given in the upper inset. The colors for the different system sizes are the same as in the Fig.~\ref{iz_thermal_shortest_con}. All quantities are dimensionless.}}
\label{iz_thermal_shortest_Sp}
\end{figure}

\textcolor{black}{In this case, to look for the quantum phase transition in the model, we consider the profiles of the order parameters (concurrence and shared purity) in the two instances for which the quantum phase transition in the one-dimensional $J_1-J_2$ model was successfully detected. These are respectively the cases $(iii)$ and $(iv)$ of Sec.~\ref{II}. Accordingly we first consider the concurrence and shared purity of a mixture of the ground and first excited states with weights as in the thermal state, and a subsequent renormalization. They are shown in Figs.~\ref{iz_thermal_shortest_con} and~\ref{iz_thermal_shortest_Sp} respectively. The plots change their curvature close to the quantum phase transition point and phase transition points of the finite-size systems shift towards the actual phase transition point for larger system sizes. The extremum points of the derivatives of the concurrence and shared purity with respect to $\lambda$, is tabulated in the tables~\ref{lambda_C_pt} and~\ref{lambda_Sp_pt} respectively.
\begin{table}[htbp]
\centering
\begin{tabular}{ |c|c|c|c|c|c| } 
 \hline
 $N$ & $10$ & $11$ & $12$ & $13$ & $14$ \\ 
 \hline
 $\lambda_{C}^{N}$ & $1.0611$ & $1.0572$ & $1.0541$ & $1.0516$ & $1.0496$ \\ 
 \hline
\end{tabular}
\caption{\textcolor{black}{System size $N$ vs maxima in the plots of derivative of the concurrence with respect to $\lambda$, for the transverse-field Ising model.}}
\label{lambda_C_pt}
\end{table}
\begin{table}[htbp]
\centering
\begin{tabular}{ |c|c|c|c|c|c| } 
 \hline
 $N$ & $10$ & $11$ & $12$ & $13$ & $14$ \\ 
 \hline
 $\lambda_{Sp}^{N}$ & $0.9652$ & $0.9712$ & $0.9774$ & $0.9801$ & $0.9819$ \\ 
 \hline
\end{tabular}
\caption{\textcolor{black}{System size $N$ vs maxima in the plots of derivative of the shared purity with respect to $\lambda$, for the transverse-field Ising model.}}
\label{lambda_Sp_pt}
\end{table}
Therefore, the order parameters applied to the reduced pseudo-thermal state detect the phase transition point. Finite-size scaling analyses for both plots are shown in the corresponding insets.
We fit a straight line
through the tabulated data points, on a log-log scale, and the scaling exponents are $0.6207$ for concurrence as the order parameter and $2.013$ for shared purity as the same. The fitted lines for concurrence and shared purity are given by Eqs.~(\ref{c_new}) and~(\ref{sp_new}), respectively :
\begin{equation}
\ln (\lambda_{C}^N-\lambda_{c})=-0.6207 \times \ln N + \mbox{constant}.
 \label{c_new}
\end{equation}
\begin{equation}
\ln (\lambda_{Sp}^N-\lambda_{c})=-2.013 \times \ln N + \mbox{constant}.
 \label{sp_new}
\end{equation}}
Note that, unlike the one-dimensional $J_1-J_2$ Heisenberg quantum spin system, the scaling exponent is same for odd- and even-site spin chains in this system.

\textcolor{black}{We then check if concurrence and shared purity of nearest neighbor density matrices of the mixture of ground and low-lying excited states as in Eq.~(\ref{density_matrix}) can detect this quantum phase transition.}   
\begin{figure}[!ht]
\includegraphics[width=\linewidth]{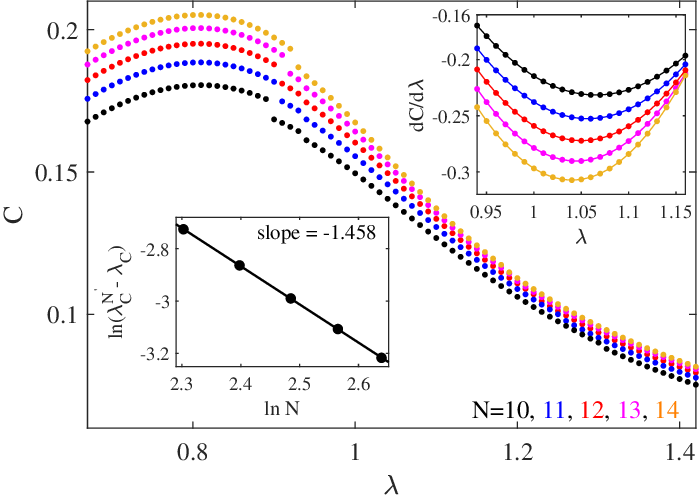}
\caption{Concurrence of the transverse-field quantum Ising model. We analyze here the concurrences of  the mixture of ground and low-lying excited states of the one-dimensional transverse-field quantum Ising model for different system sizes and their derivatives (in upper inset), and the finite-size scaling of the minimum points of the derivatives (in lower inset). All horizontal axes are dimensionless. The vertical axes of the lower inset is also dimensionless, while those of the main plot and the upper inset are in ebits.}
\label{iz}
\end{figure}

From Fig.~\ref{iz}, we note that the behavior of the concurrence in the nearest-neighbor density matrix obtained from the mixed state is similar to that of the concurrence in the nearest-neighbor density matrix of the ground state (as found in Refs.~\cite{Osterloh2002,PhysRevA.66.032110}). We plot the derivatives of the concurrence curves and fit a cubic polynomial through them in the region of interest, for different system sizes. The minimum points in the plot of derivatives indicate the corresponding quantum phase transition points - with finite-size effects 
- and are tabulated in Table~\ref{lambda}.

\begin{table}[htbp]
\centering
\begin{tabular}{ |c|c|c|c|c|c| } 
 \hline
 $N$ & $10$ & $11$ & $12$ & $13$ & $14$ \\ 
 \hline
 $\lambda_{C}^{N^{'}}$ & $1.0655$ & $1.0571$ & $1.0503$ & $1.0447$ & $1.0401$ \\ 
 \hline
\end{tabular}
\caption{System size $N$ vs minima in the plots of derivative of the concurrence with respect to $\lambda$, for the transverse-field quantum  Ising model.}
\label{lambda}
\end{table}
 We fit a straight line 
 through the tabulated data points, on a log-log scale, and find the scaling exponent to be  $-1.458$, which is weaker than the scaling exponent \(-1.87\)~\cite{Osterloh2002}, when only the ground state concurrence is used to detect the same quantum phase transition. The fitted line in the current case is given by
\begin{equation}
\ln (\lambda_{C}^{N^{'}}-\lambda_c)=-1.458 \times \ln N + \mbox{constant}.
 \label{straight_line_fit_lambda}
\end{equation}
\begin{figure}[!ht]
\includegraphics[width=\linewidth]{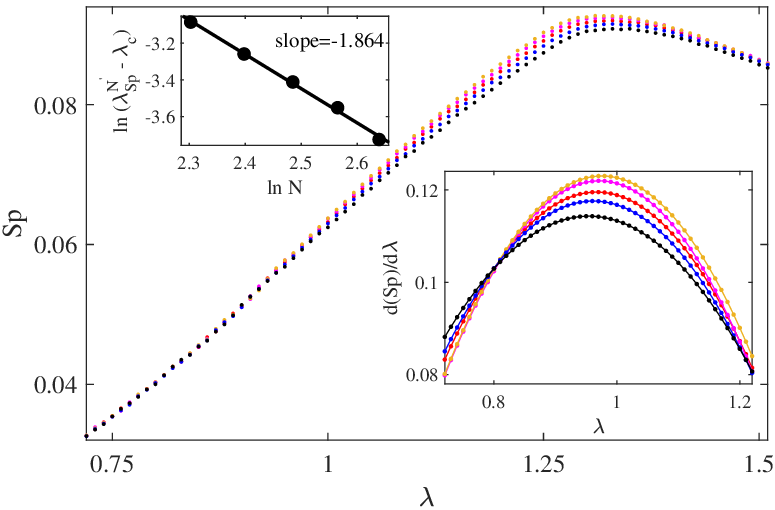}
\caption{Shared purity of the mixture of ground and low-lying excited states of one-dimensional transverse-field quantum Ising model. The different plots are for different system sizes. The derivatives are plotted in the lower inset, while the scaling analysis is given in the upper inset. The colors and symbols for the different system sizes are the same as in the preceding figure.
All quantities plotted are dimensionless.}
\label{iz_Sp}
\end{figure}

\begin{table}[htbp]
\centering
\begin{tabular}{ |c|c|c|c|c|c| } 
 \hline
 $N$ & $10$ & $11$ & $12$ & $13$ & $14$ \\ 
 \hline
 $\lambda_{Sp}^{N^{'}}$ & $0.9543$ & $0.9616$ & $0.9670$ & $0.9713$ & $0.9759$ \\ 
 \hline
\end{tabular}
\caption{System size $N$ vs maxima in the plots of derivative of the shared purity with respect to $\lambda$, for the transverse-field Ising model.}
\label{lambda_Sp}
\end{table}

We now move over to shared purity while still remaining with the transverse Ising model. 
From Fig.~\ref{iz_Sp}, we notice that the shared purity for odd- and even-site  spin chains are again similar. The general behavior of the plots for shared purity are however different in comparison with the same for concurrence, near the quantum phase transition point. In particular, in the case of concurrence, the plots show maxima at $\lambda\approx0.8$ and change their curvature from convex to concave near the quantum phase transition point, whereas in the case of shared purity, the plots show maxima at a higher value of $\lambda\approx1.25$ and change their curvature from concave to convex near the phase transition point.
The maxima in the derivatives of the shared purity versus driving parameter $\lambda$ plots indicate the phase transition points. The positions of the transition points with finite-size effects are tabulated in Table~\ref{lambda_Sp}. We fit a straight line
through the tabulated data points, on a log-log scale, and find the scaling exponent to be  $-1.864$, which is stronger as compared to the scaling exponent found in the case of concurrence. The fitted line is given by
\begin{equation}
\ln (\lambda_{Sp}^{N^{'}}-\lambda_{c})=-1.864 \times \ln N + \mbox{constant}.
 \label{straight_line_fit_lambda_Sp}
\end{equation}


\section{Two-dimensional $J_1-J_2$ Heisenberg spin model} \label{IV}

We now consider the two-dimensional $J_1-J_2$ Heisenberg  model on a  $4\times4$ square lattice, whose Hamiltonian can be written as
\begin{equation}
H_{2-D}=J_1 \sum \overrightarrow{\sigma}_i \cdot \overrightarrow{\sigma}_{j}+J_2 \sum \overrightarrow{\sigma}_i \cdot \overrightarrow{\sigma}_{k}
 \label{hamiltonian_2d}
\end{equation}
where the first sum runs over nearest neighbor pairs while the second runs over nearest neighbor diagonal pairs. 
$J_1$ and $J_2$, the coupling coefficients of nearest neighbor and diagonal neighbor spin site pairs respectively are both positive. We impose the periodic boundary condition for this two-dimensional model as well. Earlier studies with this model have predicted that there are two quantum phase transitions present - one from ordinary N\'eel order phase to a plaquette or columnar dimer phase at $\alpha \approx 0.4$ and another one from the plaquette or columnar dimer phase to a collinear N\'eel phase at $\alpha \approx 0.6$~\cite{Schulz_1992,PhysRevB.51.6151,PhysRevB.89.241104,Cysne_2015,PhysRevB.81.144410,Richter2010}.

\textcolor{black}{In this case again as in Sec.~\ref{III}, to look for the quantum phase transitions in the model, we consider the profiles of the order parameters (concurrence and shared purity) in the two instances for which the quantum phase transition in the one-dimensional $J_1-J_2$ model was successfully detected. Those are respectively the cases $(iii)$ and $(iv)$ of Sec~\ref{II}. Accordingly we first investigate the concurrence and shared purity of the reduced pseudo-thermal state with contributions from the ground and first excited states only. The results obtained have been plotted in Fig.~\ref{2d_g1}. 
\begin{figure}[!ht]
\includegraphics[width=\linewidth]{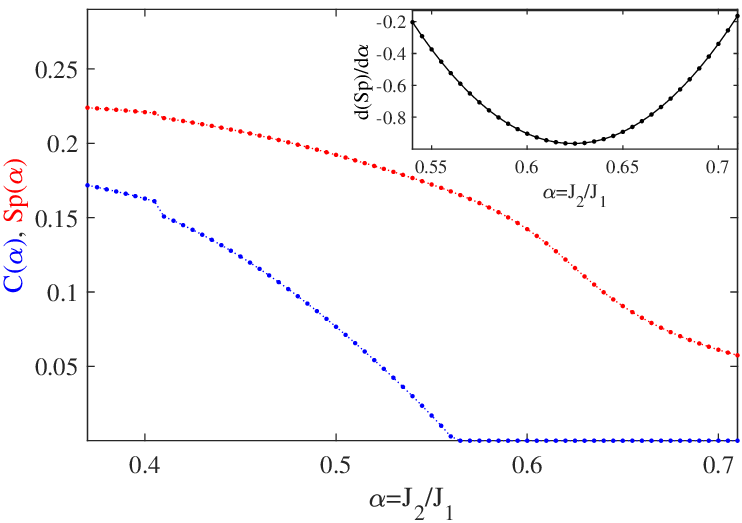}
\caption{\textcolor{black}{Shared purity and concurrence in the two-dimensional \(J_1-J_2\) spin model considering a pseudo-thermal 
state with contributions from 
the first two terms in Eq.~(\ref{thermal_state}) and the corresponding partition function. In the main figure, we plot both shared purity $(Sp)$ (red dots) and the concurrence $(C)$ (blue dots) against the system parameter \(\alpha\). The shared purity has a change of curvature near \(\alpha = 0.6\), and we fit a cubic polynomial to the data for shared purity in the vicinity of that point. The derivative  of the polynomial fit is given in the inset. 
The horizontal axes are dimensionless. The vertical axis of the inset and the vertical axis of the main figure, when used for shared purity, is also dimensionless. The vertical axis for the main figure, when used for concurrence, is in ebits.
} 
}
\label{2d_g1}
\end{figure}
We observe that the concurrence and shared purity of the mixture of the ground state and the first excited state with weights as in the thermal state can detect the quantum phase transitions of the two-dimensional $J_1-J_2$ Heisenberg spin model. The quantum phase transition point from ordinary N\'eel order phase to a plaquette or columnar dimer phase is indicated by a sharp drop at  
$\alpha^{16}_{c_{1}}\approx 0.4078$.
Bipartite or genuine multipartite entanglement of the ground state cannot detect this quantum phase transition conclusively~\cite{PhysRevA.90.032301}.
The quantum phase transition point from the plaquette or columnar dimer phase to a
collinear N\'eel phase is detected by shared purity when it changes its curvature near the quantum phase transition point. Note that concurrence does not see this phase transition. We fit a cubic polynomial in the region of interest through the data points corresponding to shared purity versus the driving parameter \(\alpha\). The derivative of the polynomial shows a minimum at $\alpha^{16}_{c_2} \approx 0.6239$, which we interpret as indicating the second quantum phase transition point.} 

\begin{figure}[!ht]
\includegraphics[width=\linewidth]{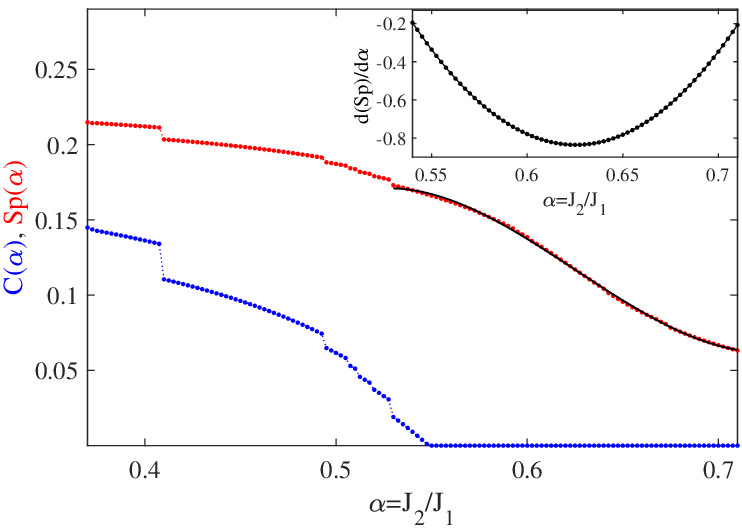}
\caption{Shared purity and concurrence in the two-dimensional \(J_1-J_2\) model considering a mixture of ground and low-lying excited states as in Eq.~\ref{density_matrix}. In the main figure, we plot both shared purity $(Sp)$ (red dots) and the concurrence $(C)$ (blue dots) against the system parameter \(\alpha\). The shared purity has a change of curvature near \(\alpha = 0.6\), and we plot  a cubic polynomial fit to the data for shared purity in the vicinity of that point, and is presented in the main figure as black line over the red dots. The derivative  of the polynomial fit is given in the inset. 
The horizontal axes are dimensionless. The vertical axis of the inset and the vertical axis of the main figure, when used for shared purity, is also dimensionless. The vertical axis for the main figure, when used for concurrence, is in ebits.
}
\label{2d}
\end{figure}

\textcolor{black}{We then investigate if concurrence and shared purity of nearest neighbor density matrices of the mixture of ground and low-lying excited states as in Eq.~(\ref{density_matrix}) can detect these quantum phase transitions.} As shown in Fig.~\ref{2d}, the concurrence as well as shared purity of the nearest-neigbor density matrix of the mixture of ground and low-lying excited states indicate the quantum phase transition from ordinary N\'eel order phase to the plaquette or columnar dimer phase by a sharp drop at $\alpha^{16}_{c_1} \approx 0.4078$.
%

The concurrence becomes zero, for the nearest-neighbor state obtained from the mixture in Eq.~(\ref{density_matrix}), after $\alpha \approx 0.55$, and cannot indicate the other quantum phase transition point (onset of the collinear N\'eel phase). However, the shared purity of the same state is non-zero in that region and changes its curvature from convex to concave, indicating the quantum phase transition from plaquette or columnar dimer phase to the collinear N\'eel order phase.
We fit a cubic polynomial in the region of interest through the data points corresponding to shared purity versus the driving parameter \(\alpha\). The derivative of the polynomial shows a minimum at $\alpha^{16}_{c_2} \approx 0.6254$, which we interpret as indicating the second quantum phase transition point.

Near $\alpha=0.5$, we notice a few discontinuities, both in concurrence and shared purity, that are not negligible compared to the first drop, which we claim to be an indication of quantum phase transition. 
The inference drawn from the presence of these discontinuities are inconclusive (to us) and may have some connection with the only phase transition in the classical limit in this model, from N\'eel to collinear at  $\alpha=0.5$~\cite{Richter2010,PhysRevA.90.032301}.

\section{Conclusions} \label{V}
\par Quantum phase transition~\cite{sachdev_2011} is a zero temperature phenomenon, but cooling a physical system to near the absolute zero temperature can be difficult. 
Hence, low-lying excited states of spin systems may form an important component of experimental realization of a physical phenomenon, especially when it is cooperative. It is therefore reasonable to consider mixtures of ground states and low-lying excited states for analyzing and characterizing cooperative physical phenomena. 

We used two physical quantities as order parameters for detecting and characterizing quantum phase transitions in certain frustrated quantum spin models as well as the non-frustrated transverse quantum Ising model. The two quantities are the shared purity and the concurrence. While the latter is a measure of entanglement, the former, although a measure of quantum correlation, is independent of the separability-entanglement paradigm. The frustrated spin models studied are the \(J_1-J_2\) Heisenberg model in one- and two- lattice dimensions. 

We found that shared purity detects the quantum phase transition point from columnar dimer to collinear N\'eel phase in the two-dimensional $J_1-J_2$ Heisenberg spin model, which is untraceable using concurrence. Furthermore, a higher finite-size scaling exponent for detection of the quantum phase transition point by shared purity in comparison to concurrence, in the case of the transverse-field Ising model, makes shared purity a promising tool for probing the connection between many-body physical systems and quantum information science. We reported different scaling exponents of quantum phase transition points for odd- and even-site spin chains governed by the one-dimensional Heisenberg $J_1-J_2$ spin model.
In the case of the one-dimensional transverse field Ising model, where no next-nearest neighbor interactions are present, the quantum phase transition points show the same scaling behavior for odd- and even-site spin chains. Such diverging behavior of scaling exponents for odd and even sized systems have been previously reported for Heisenberg ladders. 

\textcolor{black}{The concurrence and shared purity in the different models have been investigated in the reduced thermal states and certain reduced pseudo-thermal states, for the detection of quantum phase transition points in the models. We find that the order parameters, when applied to the thermal states fail to conclusively detect the phase transition points in the one dimensional \(J_1-J_2\) spin system. The same happens for certain pseudo-thermal states also. However, when the same order parameters are considered for pseudo-thermal states with only the ground and first excited states, the detection of the critical points in the one-dimensional $J_1-J_2$ model is successful. A similar success is obtained in a mixture of low-lying energy eigenstates with exponentially decaying probabilities. We subsequently utilize the states corresponding to the successful cases for detection of quantum phase transitions in the one-dimensional transverse Ising model and the two-dimensional $J_1-J_2$ model using the same order parameters. }


\acknowledgements
 US acknowledges partial support from the Department of Science and Technology, Government of India through the QuEST grant (grant number DST/ICPS/QUST/Theme-3/2019/120).

\bibliography{4th_paper_references}
\end{document}